\documentclass[prb,reprint,twocolumn,amssymb,superscriptaddress]{revtex4-2}

\usepackage{xcolor}

\usepackage{lineno}

\usepackage{bm}
\usepackage{amsmath}  
\usepackage{amsfonts}
\usepackage{graphicx}
\usepackage{hyperref}
\usepackage{comment}
\usepackage{soul}
\hypersetup{
	colorlinks=true,
	citecolor=blue,
	linkcolor=red,
	urlcolor = blue
}

\newcommand{\xvec}{\bm{x}}
\newcommand{\pvec}{\bm{p}}
\newcommand{\qvec}{\bm{q}}
\newcommand{\Hhat}{\hat{H}}

\newcommand{\grad}{\nabla}

\begin{document}

\title{Nonadiabatic Wave-Packet Dynamics: Nonadiabatic Metric, Quantum Geometry, and {\color{black}Gravitational Analogy}}

\author{Yafei Ren}
\email{yfren@udel.edu}
\affiliation{Department of Physics and Astronomy, University of Delaware, Newark, DE 19716, USA}

\author{M. E. Sanchez Barrero}
\affiliation{Department of Physics and Astronomy, University of Delaware, Newark, DE 19716, USA}

\begin{abstract}
We develop a unified theory for the nonadiabatic wave-packet dynamics of Bloch electrons subject to slowly varying spatial and temporal perturbations. Extending the conventional wave-packet ansatz to include interband contributions, we derive equations for the interband coefficients using the time-dependent variational principle, referred to as the wave-packet coefficient equation. Solving these equations and integrating out interband contributions yields the leading-order nonadiabatic corrections to the wave-packet Lagrangian. These corrections appear in three forms: (i) a nonadiabatic metric in real and momentum space, which we identify with the energy-gap-renormalized quantum metric, (ii) modified Berry connections associated with the motion of the wave-packet center, and (iii) an energy correction arising from spatial and temporal variations of the Hamiltonian. This metric reformulates the wave-packet dynamics as forced geodesic motion in phase space, enabling a gravitational analogy perspective in condensed matter systems. As an application, we analyze one-dimensional Dirac electron systems under a slowly varying exchange field $\bm{m}$. 
\end{abstract}

\maketitle

\section{Introduction}
Understanding the dynamics of Bloch electrons is crucial for gaining insight into the behavior of crystalline materials.
In the adiabatic response regime, semiclassical wave-packet theory~\cite{jones1934general, chang1995berry, chang1996berry, sundaram1999wave} provides a powerful and intuitive framework for describing electron motion. This approach, along with other related formulations~\cite{jones1934general, chang1995berry, chang1996berry, sundaram1999wave, shindou2006artificial, shindou2008gradient, wong2011quantum, wickles2013effective, bettelheim2017derivation, valet2023semiclassical, xu2025quantum}, projects the complex dynamics of many Bloch bands onto a reduced description involving only a single band or a few relevant bands. As a result, the essential physics is captured using a small number of effective parameters while eliminating irrelevant microscopic details.

The study of wave packets of Bloch electrons, pioneered by Bloch, Peierls, Jones and Zener in the early 1930s~\cite{jones1934general}, provided an intuitive foundation of understanding transport and quantum dynamics of Bloch electrons in metals and semiconductors under electromagnetic fields~\cite{slater1949electrons, luttinger1951effect, en1952motion, karplus1954hall, kohn1957quantum, luttinger1958theory, adams1959energy}. Modern developments by Chang, Sundaram, and Niu~\cite{chang1995berry, chang1996berry, sundaram1999wave} highlighted the importance of the Berry phase that is accumulated during the adiabatic time evolution~\cite{berry1984quantal, simon1983holonomy, zak1989berry} by connecting it to the anomalous velocity~\cite{karplus1954hall, adams1959energy}. This phase modifies the wave packet dynamics via introducing Berry connections $\bm{A}_{q,x,\tau}$ in its effective Lagrangian,
\begin{align}
    L_{\rm eff,a} &= \bm{q}_c \cdot \dot{\bm{x}}_c - \mathcal{E}(\bm{q}_c, \bm{x}_c) \\
    & + \dot{\bm{q}}_c \cdot \bm{A}_{q} + \dot{\bm{x}}_c \cdot \bm{A}_{x} + {\color{black}\dot{\tau} A_{\tau}}  \notag
\end{align}
where $(\bm{q}_c, \bm{x}_c)$ is the wave packet center in momentum and real spaces and $\mathcal{E}(\bm{q}_c, \bm{x}_c)$ is the energy of the wave packet.
{\color{black}The adiabatic theory highlights the emergence of Berry phase effects and the corrections to the canonical structure of the phase space and modified density of states, providing a coherent understanding of diverse Berry-phase effects~\cite{chang2008berry, xiao2010berry}. 
Corrections to the adiabatic evolution induced by electric field, magnetic field~\cite{gao2014field, gao2015geometrical}, or electric field gradient~\cite{gao2019nonreciprocal, lapa2019semiclassical, kozii2021intrinsic, liu2025geometric} have also been studied recently. These works highlighted the concept of Berry connection polarizability~\cite{gao2014field, gao2015geometrical} where those Berry connections are modified by external fields. This refinement enables descriptions of nonlinear electronic responses to external fields~\cite{yu2024quantum, liu2025quantum, verma2025quantum}, a crucial part of recent advances in higher-order novel transport phenomena beyond linear responses~\cite{jain2025anomalous, yu2024quantum, liu2025quantum, verma2025quantum, michishita2022dissipation, ahn2022riemannian}.}

Beyond the adiabatic regime, the nonadiabatic dynamics of Bloch electrons have attracted increasing attention in recent years. Advances in THz pump-probe experiments~\cite{salen2019matter}, nonlinear phononics~\cite{forst2011nonlinear, subedi2021light, levchuk2025nonlinear}, spintronics~\cite{han2023coherent}, and magnonics~\cite{pirro2021advances, bae2022exciton} now routinely drive quantum materials using time-dependent perturbations, such as electromagnetic fields, lattice vibrations, and spin precessions, at finite frequencies. Moreover, spatial inhomogeneities arising from magnetic fields, spin textures~\cite{nagaosa2019emergent, yokouchi2020emergent}, or finite wave vectors of collective excitations~\cite{nie2023surface}, play a crucial role, influencing both electronic responses~\cite{han2023coherent, nie2023surface} and their feedback on lattice~\cite{li2025nonadiabatic} and spin dynamics~\cite{lenzing2022emergent}. Nonadiabatic effects can also be induced by static strong fields, such as the Landau-Zener effects, featuring nonlinear current-field relationships~\cite{ivakhnenko2023nonadiabatic}. A unified theoretical framework that provides analytical insights into the nonadiabatic dynamics of Bloch electrons is highly desired.

Traditionally, for periodically driven systems, Floquet theory has been a powerful framework for analyzing the spectral properties, which offers analytical insights usually in the high-frequency limit~\cite{oka2019floquet, rudner2020band}. However, in the low-frequency regime, Floquet theory becomes analytically intractable and is often addressed through numerical simulations, which have motivated efforts to clarify its correspondence with adiabatic theory~\cite{rigolin2010adiabatic, rodriguez2018floquet}. It also faces substantial challenges when dealing with spatial inhomogeneities and is not applicable to nonadiabatic effects arising from strong static fields. 
To address transport properties, Floquet theory is often combined with the Keldysh nonequilibrium Green’s function formalism~\cite{sieberer2016keldysh, zhang2024achieving, mosallanejad2024floquet}. While this combined approach is powerful and offers a versatile numerical tool and serves as a valuable benchmark for strongly driven systems, the calculations are usually limited to finite-size systems and are numerically heavy, which makes it difficult to extract transparent analytical insights into the underlying physical mechanisms.

In this work, we introduce a generalized semiclassical wave-packet framework that extends beyond the adiabatic regime to capture nonadiabatic effects induced by low-frequency driving, strong static fields, and/or spatial inhomogeneity.
In our formalism, we generalize the wave function ansatz of the wave packet to include inter-band contributions and integrate out the inter-band contributions near the adiabatic regime to capture the leading-order nonadiabatic effects, which are expressed as a correction to the effective Lagrangian. We find that the nonadiabatic effects are manifested in three aspects. Firstly, a metric tensor in $(\bm{q}_c, \bm{x}_c)$-space is introduced that extends the dimension of the phase space of the wave packet. When the metric is invertible, the equation of motion can be expressed as a forced geodesic equation in $(\bm{q}_c, \bm{x}_c)$-space, an analogous of particle dynamics in a gravity field. Secondly, the Berry connections are corrected due to the interplay between wave-packet center dynamics and spatially or temporally varying Hamiltonian, generating emergent electro-magnetic fields beyond the adiabatic Berry connections. Thirdly, the energy of the wave packet is also modified by the spatial and temporal variation of the Hamiltonian. Next, we apply our formalism to study the wave packet dynamics in a one-dimensional (1D) Dirac electron system under spatially and temporally varying exchange field $\bm{m}$ as concrete examples. 

The paper is organized as follows. We present our formalism and derive the effective Lagrangian in Sec. II, study the equation of motion in Sec. III, study 1D Dirac electron systems as concrete examples in Sec. IV, and conclude with a summary in Sec. V.

\section{Formalism}
We study the dynamics of a wave packet centered at \(\xvec_c\). Its spatial spread is much smaller than the perturbation scale, which enables us to consider an approximated Hamiltonian local to the wave packet. We next construct a basis local to the wave packet and provide a detailed description of the wave packet. 

\subsection{Local Hamiltonian Approximation}
{\color{black}From the perspective of a wave packet, only the local environment within its spatial extent is accessible. When the external potential varies slowly on this scale, the Hamiltonian in the vicinity of $\bm{x}_c$ can be regarded as translationally invariant and approximated by a reference Hamiltonian $\hat{H}_c[\hat{\bm{x}},\hat{\bm{p}}]$ with $\hat{\bm{x}},\hat{\bm{p}}$ being the position and momentum operators. Translation invariance enables the definition of local Bloch states, which serve as a basis for the wave function of the wave packet. Moreover, $\hat{H}_c$ parametrically depends on $\bm{x}_c$ as well as the parameter $\tau$, denoted as $\Hhat_c = \Hhat_c[\hat{\xvec}, \hat{\pvec}; (\xvec_c, \tau)]$. Here $\tau$ encapsulates the time-dependence of the Hamiltonian. The leading-order difference between the original Hamiltonian $\hat{H}$ and $\hat{H}_c$ can be described by the gradient correction $\hat{H}_1$~\cite{sundaram1999wave} where
\begin{align}
    \Hhat \approx& \Hhat_c + \hat{H}_1, \\
    \hat{H}_1 =& \frac{1}{2} \left[  (\hat{\xvec} - \xvec_c) \cdot \grad_{\xvec_c} \Hhat(\xvec_c, \tau) + h.c. \right].
\end{align}
The spatiotemporal dependence of the Hamiltonian could be induced, for instance, by electric and magnetic fields, magnetization texture, or lattice displacements. 
For example, one can consider a wave packet moving in a crystal with slowly varying magnetization texture $\bm{m}(\bm{x},\tau)$. Given the wave packet center $\bm{x}_c$ and time-dependent parameter $\tau$, one can find the local magnetization configuration $\bm{m}(\bm{x}_c,\tau)$. We assume that $\bm{m}$ varies slowly across the scale of the wave packet. Thus we can replace the magnetization configuration with $\bm{m}(\bm{x}_c,\tau)$ to obtain a local Hamiltonian $\Hhat_c$ that has the translation symmetry of the unperturbed crystal. 
The gradient correction Hamiltonian $\hat{H}_1$ is treated perturbatively, where $\grad_{\xvec_c} \Hhat(\xvec_c, \tau)$ also has the same translation symmetry as the unperturbed crystal.}

{\color{black}
The translation invariance of the local Hamiltonian $\Hhat_c$ enables the definition of the instantaneous local Bloch bands and Bloch wave functions that depend on parameters $(\xvec_c, \tau)$:
\begin{align}
    \Hhat_c(\xvec_c, \tau) |\psi_{n,\bm{q}}(\xvec_c, \tau)\rangle = E_{c,n}(\qvec, \xvec_c, \tau) |\psi_{n,\qvec}(\xvec_c, \tau)\rangle,
\end{align}
where $\qvec$ is the Bloch wave vector, and $E_{c,n}$ denotes the $n$-th Bloch band, and $|\psi_{n,\qvec}\rangle$ is the corresponding Bloch wave function. We focus on the abelian case with nondegenerate bands. The translation invariance of $\grad_{\xvec_c} \Hhat(\xvec_c, \tau)$ enables the evaluation of the gradient correction induced intra- and inter-band couplings on the local basis.} 

\subsection{Nonadiabatic Wave Packet Ansatz}
We use the following wave packet ansatz~\cite{SM}:
\begin{align}
|\Psi(t)\rangle = \int d\bm{q} \, a(\boldsymbol{q}, t) |\tilde{\psi}_{\boldsymbol{q}}\rangle,    
\end{align}
where 
{\color{black}
$a(\bm{q},t)=|a|e^{-i\gamma}$ is a complex distribution function in the momentum space. $|a|$ is assumed to be narrow in the momentum space, enabling a well-defined wave-packet center in the momentum space
\begin{align}
    \bm{q}_c =& \int  \bm{q}|a|^2 d\bm{q} .
\end{align}
The phase $\gamma$ contributes to the wave-packet center in real space as discussed latter.
$|\tilde{\psi}_{\boldsymbol{q}}\rangle$ is a superposition of Bloch waves with wave vector $\bm{q}$
\begin{align}
|\tilde{\psi}_{\boldsymbol{q}}\rangle 
=& e^{i \boldsymbol{q} \cdot \hat{\boldsymbol{x}}} |\tilde{u}_{\bm{q}}\rangle \\
|\tilde{u}_{\bm{q}}\rangle 
=& \sum_n c_n(\boldsymbol{q}, t) |u_{n,\qvec}\rangle 
\end{align}
with $|u_{n,\qvec}\rangle$ the periodic part of the Bloch state. $c_n$ stands for the coefficient for different bands that is independent of $\xvec_c$, and $\sum_n |c_n(\boldsymbol{q}, t)|^2 = 1$. 
To ensure the invariance of the trial wave function with respect to the $U(1)$ gauge transformation in each nondegenerate band, we set the coefficients $c_n$ to satisfy the parallel-transport condition
$i\partial_{\bm q} c_n = - \bm A_{\bm q}^{nn} c_n$ with $\bm{A}_{\bm q}^{nn}=\langle u_n|i\partial_{\bm{q}} u_n\rangle$. As a result, given $c_n(\bm q_c)$, one can find $c_n({\bm q})=c_n(\bm q_c)e^{i\int_{\bm q_c}^{\bm q}\bm A_{\bm q}^{nn}\cdot d\bm q}$.
This condition ensures that $c_n$ compensates for the local phase variation of the basis states, so that the full wave function $|\Psi\rangle$ is independent of the gauge choice. 

The real-space wave-packet center is defined as 
\begin{align}
    \bm{x}_c &\doteq \langle\Psi|\hat{\bm{x}}|\Psi\rangle = \left[ \frac{\partial \gamma}{\partial \bm{q}}  + \sum_{m\neq n} c_m^* c_n \bm{A}_{\bm{q}}^{mn} \right]_{\bm{q}_c}
\end{align}
where $\bm{A}_{\bm q}^{mn}=\langle u_m|i\partial_{\bm{q}} u_n\rangle$.
%
In our nonadiabatic theory, $|c_0| \lesssim 1$ while $|c_n|\ll |c_0|$ for $n\neq0$. 
The nonadiabatic effects are manifested in the small but nonzero $c_n$. As shown later, $c_n$ are proportional to the rate of change of the dynamical parameters ($\dot{\xvec}_c$, $\dot{\qvec}_c$, and $\dot{\tau}$) as well as the inter-band coupling induced by the static gradient correction.
The nonzero $c_n$ introduces a positional shift of the wave packet center in a compact form~\cite{gao2014field}.}

\subsection{Wave Packet Coefficient Equation}
The dynamics of the wave packet is obtained by minimizing the Dirac-Frenkel action~\cite{dirac1930note, kramer1981geometry, raab2000dirac, de2025simultaneous}
\begin{align}
    \mathcal{S}=\int dt \left\langle\Psi \left|i\frac{d}{dt}-\hat{H}\right|\Psi\right\rangle
\end{align}
where $\bm{x}_c$, $\bm{q}_c$, $c_n$, and $c_n^*$ are dynamical variables. Here $d/dt$ indicates the derivative with respect to the time dependence of the wave function explicitly or implicitly through $(\qvec, \xvec, \tau, c_n)$.  
The Lagrangian is
\begin{align} \label{eq:full_lag}
    L
&=\left\langle\Psi \left|i\frac{d}{dt}-\hat{H}\right|\Psi\right\rangle \notag \\
&=\dot{\gamma}+\sum_n c_n^*i\dot{c}_n - \sum_{mn} c_m^* H^{mn} c_n \notag \\
&- \dot{\bm{q}}_c\cdot \bm{x}_c + \dot{\bm{q}}_c \cdot \tilde{\bm{A}}_q'+ \dot{\bm{x}}_c \cdot \tilde{\bm{A}}_x+ \dot{\tau} \tilde{{A}}_{\tau}
\end{align}
where the dynamic Berry connections are
\begin{align}
    \bm{\tilde{A}}_q' &= \sum_{m,n} c_m^* c_n \bm{A}_{q}^{mn} \\
    \bm{\tilde{A}}_x &= \sum_{m,n} c_m^* c_n \bm{A}_{x}^{mn} \\
    \tilde{A}_{\tau} &= \sum_{m,n} c_m^* c_n A_{\tau}^{mn} 
\end{align}
with $A_{\alpha}^{mn}=\langle u_m |i\partial_{\alpha}|u_n \rangle$ the Berry connection.
These quantities are evaluated at $(\qvec, \xvec, \tau)$. $\dot{\alpha}$ indicates the total time derivative of $\alpha$. Specifically, $\dot{c}_n = \partial_t c_n + \dot{\bm{q}}_c \partial_{\bm{q}}c_n|_{\bm{q}_c}$. 
The gradient correction-induced energy correction and interband coupling are
\begin{align}
    H^{mn} &= \langle \psi_m | \hat{H} | \psi_n \rangle = E_{c,n}\delta^{mn} + H_1^{mn} \\
    H_1^{mn} &= \langle \psi_m | \hat{H}_1 | \psi_n \rangle .
\end{align}
$H_1^{00}$ is the intra-band gradient correction that has been studied in the adiabatic case~\cite{sundaram1999wave}. The nonadiabatic corrections involve the inter-band gradient corrections $H_1^{n0}$ that are gauge covariant. 
$H_1^{0n}=(H_1^{n0})^*$ with $^*$ indicating the complex conjugate.
It is noted that, at this step, the effective Lagrangian is accurate under the assumption that \( a(\bm{q}) \) is a narrow distribution in the momentum space.

This Lagrangian gives rise to the full dynamics of the wave packet, including the evolution of its centers and the inter-band coefficients $c_n$. The equations of motion can be derived by minimizing the action $\mathcal{S}$ to $c_n^*$, $c_n$, $\bm{q}_c$, and $\bm{x}_c$. 
Specifically, the equation of motion for $c_n$ is obtained from the variational condition $\delta L/\delta c_n^* = 0$:
\begin{align}     \label{eq:c_n_motion0} 
    i\dot{c}_n &= E_{c,n} c_n - \Lambda^{nn} c_n - \sum_{m\neq n} \Lambda^{nm} c_m 
\end{align}
where $\Lambda^{nn}$ and $\Lambda^{nm}$ represent the intra- and inter-band matrix elements. Both are given by
\begin{align}
    {\Lambda}^{nm} =& \dot{\bm{q}}_c \cdot \bm{A}_q^{nm} + \dot{\bm{x}}_c \cdot \bm{A}_x^{nm} + \dot{\tau} A_\tau^{nm} - H_1^{nm}.
\end{align}
The terms in $\Lambda^{nm}$ other than $H_1^{nm}$ distinguish Eq.\eqref{eq:c_n_motion0} from the standard Schr\"odinger equation governing the time evolution of the coefficients for a set of fixed basis. By choosing the local instantaneous eigenstates of $\hat{H}_c$ as the basis, the evolution of the coefficients $c_n$ is determined not only by the spatiotemporal dependence of the local Hamiltonian, but also by the dynamics of the wave packet centers $\bm{\xi} = ({\bm{q}}_c, {\bm{x}}_c)$.
We therefore term Eq.~\eqref{eq:c_n_motion0} the wave packet coefficient equation.

For later convenience, we further decompose $\Lambda^{nm}$ as
\begin{align} \label{eq:Lambda_decompose}
    {\Lambda}^{nm} =& \, \dot{\bm{\xi}}\cdot \bm{A}_{\xi}^{nm} + \bar{\Lambda}^{nm}, \\
    \bar{\Lambda}^{nm}=& \, \dot{\tau} A_\tau^{nm} - H_1^{nm} \label{eq:Lambda_decompose1}
\end{align}
where $\bar{\Lambda}^{nm}$ captures the effects arising from the explicit temporal variation and spatial gradient of the Hamiltonian, while the term $\dot{\bm{\xi}} \cdot \bm{A}_{\xi}^{nm}$ accounts for the intraband corrections and interband transitions induced by the motion of the wave packet centers.

\subsection{Effective Lagrangian}
We derive the effective Lagrangian of the wave packet center $\bm{\xi}$ by integrating out the dynamics of $c_n$. By using the saddle point elimination method, we solve the equations of motion of $c_n$, express them as a function of $\bm{\xi}$ and their time derivatives, and substitute $c_n$ back into the Lagrangian $L$. This process enables us to decouple the dynamics of $(\qvec_c, \xvec_c)$ from $c_n$, leading to an effective Lagrangian of the wave packet centers. Specifically, we find that the leading order corrections can be expressed as~\cite{ren2025momentum}
\begin{align}
c_n(t) \simeq \frac{{\Lambda}^{n0}(t)}{ \omega^{n0}} c_0(t)
\end{align}
where the energy gap $\omega^{n0}=E_{c,n}-E_{c,0}$.
Given that $|c_0|^2=1-\sum_{n\neq0} |c_n|^2\simeq 1$, we find that, for $n\neq 0$,
\begin{align}
    c_0^*c_n\simeq \frac{{\Lambda}^{n0}(t)}{ \omega^{n0}}.
\end{align}
Substituting this back into Eq.~\eqref{eq:full_lag} and making use of $|c_0|^2=1-\sum_{n\neq0}|c_n|^2$, we obtain the effective Lagrangian
\begin{align}
    L_{\rm eff} 
    = L_{\rm eff,a} + L_{\rm eff,na}.
\end{align}
Here $L_{\rm eff,a}$ is the adiabatic effective Lagrangian 
\begin{align}
    L_{\rm eff,a} &= - \dot{\bm{q}}_c \cdot \bm{x}_c - E_{c,0} + \Lambda^{00} 
\end{align}
where $\Lambda^{00}$ includes Berry connection corrections and $H_1^{00}$ that is the gradient correction of the energy of the $0$-th band.

$L_{\rm eff,na}$ is the leading-order nonadiabatic correction that has a compact form 
\begin{align}\label{eq:Leffna}
    L_{\rm eff,na} = \sum_{n\neq 0} \frac{|{\Lambda}^{n0}|^2}{ \omega^{n0}}.
\end{align}
$L_{\rm eff,na}$ is gauge invariant following the gauge covariance of ${\Lambda}^{n0}$. By using Eq.~\eqref{eq:Lambda_decompose}, the nonadiabatic correction becomes
\begin{align}
    L_{\rm eff,na} = \frac{1}{2}\sum_{ij} G_{ij}\dot{\xi}^i \dot{\xi}^j  + \sum_{i} \dot{\xi}^i \delta A_i + \delta E_{\rm na}
\end{align}
where 
\begin{align}
    G_{ij} =& \, 2{\rm Re}\sum_{n\neq 0}\frac{A_{i}^{0n}A_{j}^{n0}}{\omega^{n0}} \\
    \delta A_i =& \, \sum_{n\neq 0}\frac{A_{i}^{0n}\bar{\Lambda}^{n0}+\bar{\Lambda}^{0n}A_{i}^{n0}}{\omega^{n0}}  \\
    \delta E_{\rm na} =& \, \sum_{n\neq 0} \frac{\bar{\Lambda}^{0n}\bar{\Lambda}^{n0}}{\omega^{n0}} .
\end{align}
Here $A_{i}^{n0}$ is the interband Berry connection for parameter $\xi^i$.
Depending on the degree of $\dot{\bm{\xi}}$, the nonadiabatic corrections modify the wave packet dynamics in three ways. The quadratic terms introduce a metric tensor $G_{ij}$ in $\bm{\xi}$-space, dubbed the nonadiabatic metric. The linear terms modify the Berry connections by $\delta {A}_i$, whereas the $\delta E_{\rm na}$ stands for the nonadiabatic energy correction, which is a second-order correction.

\subsubsection{Nonadiabatic Metric $G_{ij}$ vs Quantum Metric}
The key distinction between the adiabatic and nonadiabatic wave packet dynamics is the $G_{ij}$ term that introduces a metric tensor in the $\xi$-space. 
As this metric originates from nonadiabatic corrections, we term it the nonadiabatic metric.
The metric is a gauge-invariant symmetric tensor, which is related to but different from the quantum metric (the Fubini-Study metric)
\begin{align}
    g_{ij} =& \text{Re} \mathcal{Q}_{ij} 
\end{align}
that is the real part of the quantum geometric tensor \begin{align}
    \mathcal{Q}_{ij}=& \sum_{n\neq 0} {A_{i}^{0n} A_{j}^{n0}}.
\end{align}
The energy denominator makes $G_{ij}$ not necessarily positive-semidefinite and thus not necessarily a Riemannian metric~\cite{nakahara2018geometry}. In a two-band model with an energy gap ${\color{black}2}\Delta(\bm{q})$, $G_{ij}$ of the lower band is proportional to $g_{ij}$: $G_{ij} = \frac{1}{\Delta}g_{ij}$, which is positive-semidefinite. For the upper band, the nonadiabatic metric is negative-semidefinite. 

The $q$-block of $G_{ij}$ makes the momentum space a curved manifold~\cite{ren2025momentum}, whereas the $x$-block makes the real space curved. There can also be a metric between $q$ and $x$ space, making $q$ and $x$ on equal footing. This is in contrast to the adiabatic theory, where the canonical momentum of $\bm{x}_c$ is a function of $\bm{q}_c$. Here $\bm{\xi}$ is the generalized coordinate and $\dot{\bm{\xi}}$ is the generalized velocity. A canonical structure is well defined when the inverse of the nonadiabatic metric $G$ exists. {\color{black}The nonadiabatic metric directly enters the dynamics of Bloch-electron wave packets, introducing additional velocities and emergent gauge fields as discussed in Sec.~\ref{sec:EoM}.}

\subsubsection{Modified Berry Connections {and Emergent Gauge Fields}}\label{sec:ModifiedA}
Berry connections have far-reaching manifestations in different aspects of the electronic properties, including density of states, polarization, magnetization, and anomalous transport~\cite{xiao2010berry}. By modifying the Berry connections, the nonadiabatic effects introduce corrections to those broad processes. 
Since $\bar{\Lambda}^{n0}$ in Eq.~\eqref{eq:Lambda_decompose1} has two contributions from the temporal variation and spatial gradient of the Hamiltonian, the modification to the Berry connection also has two contributions 
\begin{align}\label{eq:deltaAi}
    \delta A_i = \dot{\tau} \delta A_{i,\tau} + \delta A_{i,1}.
\end{align}

The dynamic correction proportional to $\dot{\tau}$ is the energy-gap weighted quantum metric in the parameter space of $\xi^i$ and $\tau$ 
\begin{align}
    \delta A_{i,\tau} =& \sum_{n\neq 0} \frac{A_{i}^{0n}A_{\tau}^{n0}+c.c.}{\omega^{n0}}.
\end{align}
It originates from the dynamical variation of the Hamiltonian induced by either time-dependent electromagnetic fields or coherent excitations, such as coherent phonons or magnons. The momentum-space Berry connection is closely related to the electric polarization and anomalous velocity. Thus, the dynamically varying Hamiltonian indicates a correction to the polarization and transport that is proportional to $\dot{\tau}$. For instance, when $\tau$ corresponds to the phonon coordinate, an oscillating dipole proportional to the phonon velocity instead of the phonon displacement can be induced.

The second term is 
\begin{align}
    \delta A_{i,1} =& \, -\sum_{n\neq 0} \frac{A_{i}^{0n}H_{1}^{n0}+c.c.}{\omega^{n0}}.
\end{align}
It originates from the interband mixing induced by the gradient correction, which is a static potential. Examples of the gradient correction are static magnetic field, electric field gradient, spin textures, and lattice dislocations. This term is closely tied to the electric polarization, electric transport, and emergent magnetic fields induced by those inhomogeneities. 

The presence of $\delta A_i$ naturally modifies the Berry curvature, which will influence the phase-space density of states and anomalous transport. Nevertheless, the Chern number is not influenced because $\delta A_i$ are gauge invariant and periodic functions in the momentum space, and the integral of their derivatives with respect to the momentum over the Brillouin zone vanishes. The real-space Berry connection plays the role of an effective vector potential of electrons. The curl and temporal variation of the vector potential generate effective magnetic and electric fields. Thus, this term indicates a way to generate emergent electro-magnetic fields using dynamically varying parameters. {\color{black}This generalizes the emergent electromagnetic induction in helical-spin magnets~\cite{nagaosa2019emergent, yokouchi2020emergent} into a more general setting. More details are presented in Secs.~\ref{subsec:geometricvelocity} and~\ref{subsec:geometricforce}.}

\subsubsection{Nonadiabatic Energy Correction and Controlling External Dynamical Parameters}\label{sec:NonadiabaticEnergy}
The nonadiabatic energy correction reads
\begin{align}
    \delta E_{\rm na} =& {\color{black}-} \delta E_{\rm na}' + \dot{\tau}\delta A_\tau 
\end{align}
where
\begin{align}
    \delta E_{\rm na}' =&\, {\color{black}-} \sum_{n\neq 0} \dot{\tau}^2 \frac{A_{\tau}^{0n}A_{\tau}^{n0}}{\omega^{n0}} + \frac{H_{1}^{0n}H_1^{n0}}{\omega^{n0}} \\
    \delta A_\tau=& \, -\sum_{n\neq 0}  \frac{A_{\tau}^{0n}H_1^{n0}+H_1^{0n}A_{\tau}^{n0}}{\omega^{n0}}.
\end{align}
The first term is proportional to $\dot{\tau}^2$, and the coefficient is also an energy-gap weighted quantum metric. The metric is defined in the parameter space spanned by $\tau$ or the parameters associated with it such as lattice displacement or spin orientations. This term not only influences the wave packet dynamics as an energy correction but also influences the dynamics of these dynamical variables. For example, this term induces an effective metric tensor or inertia in the dynamics of lattice or spin. This metric tensor in $\tau$-space is thus dubbed the nonadiabatic $\tau$-metric and will be discussed in a separate work. 
The second term in $\delta E_{\rm na}'$ is the second-order energy correction induced by the gradient corrections. It is thus proportional to the second order of the inhomogeneities induced by, e.g., the magnetic field or electric field gradient.

The $\delta A_{\tau}$ appears in the presence of both temporal variation and the gradient correction. It introduces a gauge-invariant correction to the Berry connection $A_{\tau}^{00}$. This enables the study of how the gradient correction modifies the dynamics of the variables associated with $\tau$, such as the influence of the magnetic field on the lattice dynamics and phonon properties.

\section{Equation of motion}\label{sec:EoM}
{\color{black}For brevity, we drop the subscript $c$ in $(\bm{q}_c, \bm{x}_c)$ from now on and summarize the effective Lagrangian below with the Einstein summation convention implied:
\begin{align}
    L_{\rm eff} = \frac{1}{2}G_{ij}\dot{\xi}^i\dot{\xi}^j + \dot{\xi}^i \left( \frac{1}{2}  J_{ij}  {\xi}^j + \bar{A}_i \right)  - \left( E-\dot{\tau}\bar{A}_{\tau} \right) 
\end{align}
where $\bar{A}_j = {A}_j + \delta {A}_j$ with $A_j$ the short-hand notation for $A_{j}^{00}$ and $\delta A_j$ the nonadiabatic correction.} Here, $\bm{J}$ is the standard symplectic matrix
\begin{align}
\bm{J} = \begin{bmatrix}
0 & -\mathbf{I} \\
\mathbf{I} & 0
\end{bmatrix}
\end{align}
with $\mathbf{I}$ the identity matrix. This term is equivalent to $- \dot{\bm{q}}_c \cdot \bm{x}_c$ up to a total time derivative.
The wave packet energy is
\begin{align}\label{eq:etotal}
    E =E_{c,0}  + H_1^{00} + \delta E_{\rm na}'
\end{align}
including the band energy $E_{c,0}$, the adiabatic gradient correction $H_1^{00}$, and the nonadiabatic energy correction $\delta E_{\rm na}'$.

The Euler-Lagrange equations for the generalized coordinates \( \xi^k \) are:
\begin{align}
    \frac{d}{dt} \left( \frac{\partial L_{\rm eff}}{\partial \dot{\xi}^k } \right) - \partial_k L_{\rm eff} = 0
\end{align}
The equations of motion are:
\begin{align} \label{eq:geodesic0}
    G_{kj} \ddot{\xi}^j + \Gamma_{k,lm} \dot{\xi}^l \dot{\xi}^m + (J_{kj} - \bar{F}_{kj}) \dot{\xi}^j +\bar{F}_{\tau k} + \partial_k E = 0
\end{align}
where the Christoffel symbol of the first kind is
\begin{align}
    \Gamma_{k,lm} = \frac{1}{2} \left( \partial_m G_{kl} + \partial_l G_{km} - \partial_k G_{lm} \right),
\end{align}
and the modified Berry curvatures are
\begin{align}
    \bar{F}_{kj} &= \partial_k \bar{A}_j - \partial_j \bar{A}_k \\
    \bar{F}_{\tau k} &= \partial_{\tau} \bar{A}_k - \partial_k \bar{A}_{\tau} .
\end{align}

\subsection{Geometric and Geodesic Velocities}\label{subsec:geometricvelocity}
{\color{black}The equation of motion in Eq.~\eqref{eq:geodesic0} contains both the dynamics of $\bm{x}$ and $\bm{q}$, representing the velocities of the wave packet and its motion in $\bm{q}$-space induced by external driving forces, separately. To explicitly show the dynamics of $\bm{x}$ and $\bm{q}$, we introduce the equation of motion of their components with $(\alpha,\beta,\gamma)$ indicating the Cartesian coordinate components. Below, we set $q^{\alpha}=\delta^{\alpha\beta} q_{\beta}$ and $x^{\alpha}=\delta^{\alpha\beta} x_{\beta}$ with $\delta^{\alpha\beta}$ the Kronecker delta. }

{\color{black} For components of $\bm{x}$, Eq.~\eqref{eq:geodesic0} can be rewritten as the following equation that defines the velocity 
\begin{align} \label{eq:xdot}
    \dot{x}_{\alpha} 
    &= \partial_{q^{\alpha}} E - \bar{{F}}_{q^{\alpha}q^{\beta}} \dot{{q}}^{\beta} - \bar{{F}}_{q^{\alpha}x^{\beta}} \dot{{x}}^{\beta} - \bar{{F}}_{q^{\alpha}\tau } {\color{black}\dot{\tau}} \notag \\
    &+ {G}_{q^{\alpha}q^{\beta}} \ddot{{q}}^{\beta} + {G}_{q^{\alpha}x^{\beta}} \ddot{{x}}^{\beta}  \\
    &+ {\Gamma}_{q^{\alpha}, q^{\beta}q^{\gamma}}\dot{q}^{\beta}\dot{q}^{\gamma}
    + 2{\Gamma}_{q^{\alpha}, q^{\beta}x^{\gamma}}\dot{q}^{\beta}\dot{x}^{\gamma}
    + {\Gamma}_{q^{\alpha}, x^{\beta}x^{\gamma}}\dot{x}^{\beta}\dot{x}^{\gamma}.\notag
\end{align}
The velocity contains multiple contributions. The group velocity is $\partial_{q^{\alpha}} E$ where $E$ in Eq.~\eqref{eq:etotal} contains leading-order gradient correction and next-order nonadiabatic corrections. $\bar{F}_{q^{\alpha} q^{\beta}}$ is the corrected Berry curvature tensor in the momentum space, which is antisymmetric with respect to exchanging $(\alpha,\beta)$, contributing to an anomalous velocity perpendicular to $\dot{\bm{q}}$. $\bar{F}_{q^{\alpha}x^{\beta}}$ arises from the fact that during the evolution of the wave packet, the accumulated Berry phase depends simultaneously on the real- and momentum-space coordinates. This term can be moved to the left-hand side, contributing to the modification of the density of states~\cite{xiao2005berry, xiao2010berry}. $\bar{F}_{q^{\alpha}\tau} \dot{\tau}$ arises from the winding of the instantaneous eigenstates in time and momentum space. For example, when the time dependence arises due to the atomic displacement $Q(\tau)$, the oscillating displacement can introduce a velocity correction $\sim \bar{F}_{q^{\alpha}\tau} \dot{\tau} \sim \bar{F}_{q^{\alpha}Q} \partial_{\tau}{Q}\dot{\tau}$, contributing to a polarization current.
These Berry curvatures have incorporated the corrections from the modified Berry connections $\delta A_i$ due to the time dependence of the Hamiltonian and the gradient corrections as discussed in Sec.~\ref{sec:ModifiedA} and Sec.~\ref{sec:NonadiabaticEnergy}}. These terms are quantitative corrections to the adiabatic wave packet dynamics~\cite{sundaram1999wave}.
{\color{black}The qualitative different contributions from the nonadiabatic dynamics is manifested in the second and third lines of Eq.~\eqref{eq:xdot}.} 

{\color{black}The velocity corrections in the second line are termed the geometric velocity, as the coefficients ${G}_{q^{\alpha}q^{\beta}}$ and ${G}_{q^{\alpha}x^{\beta}}$ are the nonadiabatic metric that describes the phase-space geometry.
These velocity corrections are proportional to the accelerations, $\ddot{q}^{\beta}$ and $\ddot{x}^{\beta}$. As the leading order contribution to the $\dot{q}^{\alpha}$ is the external driving force, such as an electric field, the geometric velocity from $\ddot{q}^{\beta}$ can introduce a component proportional to the frequency of the driving force, which is inherently important in the frequency-dependent electronic responses. 
The $\ddot{x}_{\alpha}$ term can be understood iteratively by calculating the time derivative of $\dot{x}_{\alpha} 
\simeq \partial_{q^{\alpha}} E - \bar{{F}}_{q^{\alpha}q^{\beta}} \dot{{q}}^{\beta} - \bar{{F}}_{q^{\alpha}x^{\beta}} \dot{{x}}^{\beta} - \bar{{F}}_{q^{\alpha}\tau } \dot{\tau}$. The leading order contribution to $\dot{x}_{\alpha}$ is the group velocity $\partial_{q^{\alpha}} E$ in general energy bands and thus $\ddot{x}_{\alpha}\sim \dot{q}^{\beta} \partial_{q^{\beta}} \partial_{q^{\alpha}} E + \dot{x}^{\beta} \partial_{x^{\beta}} \partial_{q^{\alpha}} E$. The first term involves the inverse of the effective mass tensor that is large in systems with small effective mass. The second term is proportional to the 
spatial gradient of the group velocity times the group velocity.}

The third line corresponds to the geodesic velocity as the coefficients ${\Gamma}_{q^{\alpha},lm}$ are Christoffel symbols of the first kind that are essential for the geodesic equation discussed below. The geodesic velocity is proportional to the quadratic products of $(\dot{x}^{\alpha}, \dot{q}^{\beta})$. The $\dot{q}^{\beta} \dot{q}^{\gamma}$ term is proportional to the quadratic order of the external driving forces, contributing to the nonlinear responses of Bloch electrons to electric and magnetic fields~\cite{yu2024quantum, liu2025quantum, verma2025quantum}. The coefficient 2 in the $\dot{q}^{\beta} \dot{x}^{\gamma}$ term originates from the fact that ${\Gamma}_{k,lm}$ is symmetric with respect to exchanging $lm$. This term is determined mainly by the product of the external driving force and the velocity, which may contribute to a spatial inhomogeneity-induced nonlinear current. The $\dot{x}^{\beta}\dot{x}^{\gamma}$ term contributes to a velocity that is proportional to the group velocity squared. The physical meaning of the latter two terms in the geodesic velocity remains poorly understood, which we leave for future study.

\subsection{Geometric and Geodesic Forces}\label{subsec:geometricforce}
The force inducing $\dot{\bm{q}}$ becomes
\begin{align}\label{eq:qdot}
\dot{q}_{\alpha} 
    &= -\partial_{x^{\alpha}} E + \bar{{F}}_{x^{\alpha}x^{\beta}} \dot{{x}}^{\beta} + \bar{{F}}_{x^{\alpha}q^{\beta}} \dot{{q}}^{\beta} + \bar{{F}}_{x^{\alpha}\tau } \dot{\tau} \notag \\
    &- {G}_{x^{\alpha}x^{\beta}} \ddot{{x}}^{\beta} - {G}_{x^{\alpha}q^{\beta}} \ddot{{q}}^{\beta}  \\
    &- {\Gamma}_{x^{\alpha}, q^{\beta}q^{\gamma}}\dot{q}^{\beta}\dot{q}^{\gamma}
    - 2{\Gamma}_{x^{\alpha}, q^{\beta}x^{\gamma}}\dot{q}^{\beta}\dot{x}^{\gamma}
    - {\Gamma}_{x^{\alpha}, x^{\beta}x^{\gamma}}\dot{x}^{\beta}\dot{x}^{\gamma}.\notag
\end{align}
{\color{black}The forces arise from spatially varying energy and emergent fields from spatial inhomogeneity.
The first term on the right-hand side stands for the force from the spatially varying total energy $E$ of the wave packet as defined in Eq.~\eqref{eq:etotal}, including the effects from the external electric field and the nonadiabatic energy corrections from spatial inhomogeneity. 
$\bar{{F}}_{x^{\alpha}x^{\beta}}$ is an antisymmetric tensor that plays the role of a real-space magnetic field. 
This term can be induced by a genuine external magnetic field or an emergent magnetic field from spin textures in the adiabatic limit~\cite{ye1999berry}. Nonadiabatic effects can also contribute to an emergent gauge field by inducing $\delta A_{x^{\alpha}}$ defined in Eq.~\eqref{eq:deltaAi}, which can be induced by spatial inhomogeneity with or without temporal driving. The $\bar{{F}}_{x^{\alpha}q^{\beta}}$ term can be moved to the left-hand side, contributing to the modified density of states~\cite{xiao2010berry}. The nonadiabatic corrections to the density of states have been studied in very recent works~\cite{mameda2025quantum, maranzana2026semiclassical}. $\bar{{F}}_{x^{\alpha}\tau}$ corresponds to an extra geometric force, arising from the non-trivial spatiotemporal winding of the instantaneous eigenstates. It captures the interplay between spatial inhomogeneity and temporal driving.}

{\color{black}The second and third lines arise from the spatially varying wave functions, which are qualitatively different from the adiabatic effects. The second line denotes forces from ${G}_{x^{\alpha}x^{\beta}}$ and ${G}_{x^{\alpha}q^{\beta}}$ are termed the geometric forces, as ${G}_{x^{\alpha}x^{\beta},x^{\alpha}q^{\beta}}$ are components in the nonadiabatic metric. The phase-space accelerations, $\ddot{x}^{\beta}$ and $\ddot{q}^{\beta}$, correspond to the temporal variation of the velocity and external driving force as discussed in the last subsection. 
The third line corresponds to the geodesic forces where the coefficients are Christoffel symbols ${\Gamma}_{x^{\alpha},\xi^i\xi^j}$. Those forces are nonlinear in external fields that drive the motion of $\bm{q}$ or nonlinear in the spatial inhomogeneity.} 

\subsection{{\color{black}Gravitational Analogy} in Phase Space}

When the metric $G_{ij}$ is invertible, {\color{black}i.e., $G^{jk}$ exists such that $G_{ij} G^{jk} = \delta_i^k$ with $\delta_i^k$ the Kronecker delta, the wave packet's equation of motion can be written in the form of a geodesic equation subject to external forces on the phase space spanned by $\bm{\xi}=(\bm{q}, \bm{x})$}:
\begin{align} \label{eq:analogousgravity}
    \ddot{\xi}^i + \Gamma^i_{jk} \dot{\xi}^j \dot{\xi}^k + G^{ik} \left[(J_{kj} - \bar{F}_{kj}) \dot{\xi}^j + \bar{F}_{\tau k} + \partial_k E \right] = 0
\end{align}
where $\Gamma^i_{jk}=G^{il}\Gamma_{l,jk}$ are the Christoffel symbols {\color{black}of the second kind} associated with the metric $G_{ij}$. This equation is mathematically analogous to the dynamics of a particle moving on a curved manifold parametrized by $\bm{\xi}$, endowed with metric $G_{ij}$, and subject to external forces. Here, the symplectic matrix $\bm{J}$ together with the Berry curvature $\bar{F}_{kj}$ acts as an effective magnetic field; $E$ plays the role of a scalar potential; and $\bar{F}_{\tau k}$ represents an additional force arising from spatio-temporal variations in the system.

{\color{black}Equation~\eqref{eq:analogousgravity} provides a useful gravitational analogy in a restricted sense: in both cases, the metric enters the equations of motion through the Christoffel symbols and geometrically shapes the inertial part of the trajectory. Unlike general relativity, however, the metric here is defined on the wave-packet phase space $(\bm q,\bm x)$, rather than on spacetime. This makes it different from the conventional study in the field of ``analogue gravity'' that employs an analogue spacetime metric~\cite{barcelo2011analogue} in bosonic~\cite{munoz2019observation, viermann2022quantum, vsvanvcara2024rotating, falque2025polariton} or fermionic~\cite{giovanazzi2005hawking, volovik2016black, fulgado2023fermion, haller2023black} systems. Moreover, $G_{ij}$ here does not obey an independent dynamical field equation. Instead, it is a static effective metric, not necessarily Riemannian or Lorentzian, determined by the underlying electronic wave function and band structure. This metric governs only the dynamics of quasielectrons and is not universal for all excitations.}

In addition, we note that the term ``emergent gravity'' is used to describe the nonadiabatic electron dynamics in a spin texture~\cite{onishi2025emergent}, though, in quantum gravity research, the same term is used to indicate scenarios where the metric dynamics emerges as a low-energy effective phenomenon from some more fundamental, non-gravitational theory~\cite{verlinde2011origin, verlinde2017emergent}. 
Similar studies have been reported in recent independent works~\cite{yoshida2025emergent, maranzana2026semiclassical}.
The analogy here is also distinct from the so-called artificial gravity explored in a parallel line of research, which focuses on electronic dynamics under spatially and temporally varying lattice deformations~\cite{dong2018geometrodynamics, li2023geodynamics} or curved surface~\cite{jiang2022geometric}.

\section{A Case Study: 1D Dirac Electron}
We consider a 1D model described by the Hamiltonian
\begin{align}\label{eq:ham1d}
    \hat{H}(q,x,\tau) = vq\,\sigma_z + g\,\bm{m}(x,\tau)\cdot\bm{\sigma},
\end{align}
where $v$ is the Fermi velocity, $q$ is the crystal momentum, $\bm{\sigma}$ denotes the Pauli matrices, $g$ is the coupling constant, and $\bm{m}(x,\tau)=(m_x,m_y,0)$ is an exchange field that varies slowly in position $x$ and time $\tau$. By ``slowly'' we mean that the spatial modulation wavelength is much larger than the lattice constant and the temporal modulation frequency is much smaller than the band gap. The local energy spectrum consists of two bands,
\begin{align}
    E_0(q,x,\tau) = -\Delta, \qquad E_1(q,x,\tau) = \Delta,
\end{align}
where $\Delta = \sqrt{(vq)^2 + (g m)^2}$ and $m = \sqrt{m_x^2 + m_y^2}$.

The adiabatic dynamics of Bloch wave packets are governed by the intra-band Berry connections $A_\alpha$ with $\alpha = q, x, \tau$ and by the gradient correction $H_1^{nn}$. For the lower band one obtains
\begin{align}
    \Lambda^{00} = \dot{q} A_q + \dot{x} A_x + \dot{\tau} A_\tau - H_1^{00}.
\end{align}
With a suitable gauge choice $A_q$ vanishes, and the remaining terms depend on the transverse, i.e., rotational, variation of $\bm{m}$:
\begin{align}
    A_x &= - \left(1 + \frac{vq}{\Delta}\right) \frac{[\bm{m} \times \partial_x \bm{m}]_z}{m^2}, \\
    A_\tau &= - \left(1 + \frac{vq}{\Delta}\right) \frac{[\bm{m} \times \dot{\bm{m}}]_z}{m^2}, \\
    H_1^{00} &= \textcolor{black}{-vg^2 \frac{[\bm{m} \times \partial_x{\bm{m}}]_z}{\Delta^2}},
\end{align}
where $\dot{\bm{m}} \equiv \dot{\tau} \partial_\tau \bm{m} $. The term $\bm{m} \times \partial_x \bm{m}$ encodes the spatial texture of $\bm{m}$ and contributes to both $A_x$ and $H_1^{00}$, while $\bm{m} \times \dot{\bm{m}}$ describes its angular velocity and enters $A_\tau$.

Beyond the adiabatic regime, nonadiabatic corrections arise from interband transitions encoded in $\Lambda^{n0}$, which depends on both the interband Berry connections and the corresponding gradient corrections. The coupling between the two bands takes the form
\begin{align}
    \Lambda^{10} = \dot{q} A_q^{10} + \dot{x} A_x^{10} + \dot{\tau} A_\tau^{10} - H_1^{10},
\end{align}
where
\begin{align}
    A_q^{10} &= \frac{i v g m}{2 \Delta^2}, \\
    A_x^{10} &= \frac{g}{2 m \Delta^2} \left[\Delta(\partial_x\bm{m} \times \bm{m})_z - i v q \partial_x\bm{m} \cdot \bm{m}\right], \\
    A_\tau^{10} &= \frac{g}{2 m \Delta^2} \left[\Delta(\dot{\bm{m}} \times \bm{m})_z - i v q \dot{\bm{m}} \cdot \bm{m}\right], \\
    H_1^{10} &= \textcolor{black}{i \frac{v g^3 m}{2 \Delta^3} \bm{m} \cdot \partial_x\bm{m}}.
\end{align}
A key observation is the appearance of terms involving $\bm{m} \cdot \partial_\alpha \bm{m} = \tfrac{1}{2} \partial_\alpha |\bm{m}|^2$, which quantify the longitudinal variation of $\bm{m}$ along $\alpha$. These longitudinal components contribute to the imaginary parts of the Berry connections and gradient corrections and thus play an important role in the nonadiabatic electron dynamics. Spatial amplitude variations enter through $A_x^{10}$ and influence the nonadiabatic metric components $G_{xx}$, $G_{xq}$, and $G_{qx}$. Temporal and spatial amplitude variations also appear separately in $A_\tau^{10}$ and $H_1^{10}$, which modify the Berry connections $\delta A_i$. Here $\delta A_q$ is associated with electric polarization while $\delta A_x$ acts as an effective vector potential. Furthermore, $A_\tau^{10}$ and $H_1^{10}$ contribute directly to energy corrections, and their cross terms further modify $A_\tau$, thereby influencing the charge pumping processes. 

\subsection{Uniform Charge Pumping}
The leading-order nonadiabatic corrections studied here has negligible effects on the quantized Thouless pumping. As an example, we set $m_x=m\cos(\omega \tau)$ and $m_y=m\sin(\omega \tau)$ with $\tau =t$. In this case, 
\begin{align}
    \Lambda^{00} =&  A_{\tau} \\
    \Lambda^{10} =& \, \dot{q}\,A_q^{10} + A_\tau^{10} 
\end{align}
where $A_{\tau}=-\omega(1+\frac{vq}{\Delta})$, $A_q^{10}=\frac{i v g m}{2 \Delta^2}$, and $A_\tau^{10}=-\frac{gm\omega}{2\Delta}$. As a result, the effective Lagrangian reads:
\begin{align}
    L_{\rm eff} 
    &= - \dot{{q}}  {x} - E_{0}(q) + A_{\tau} + \frac{1}{2}G_{qq}\dot{q}^2+ \delta E_{\rm na}'
\end{align}
where $G_{qq}=2\frac{|A_q^{10}|^2}{{\color{black}2}\Delta}$ and $\delta E_{\rm na}'= {\color{black}-}\frac{|A_\tau^{10}|^2}{{\color{black}2}\Delta}$. The corresponding equation of motion reads
\begin{align}
    \dot{x}=&\partial_q(E_0{\color{black}+}\delta E_{\rm na}') + \partial_q A_{\tau} + G_{qq}\ddot{q} + \Gamma_{q,qq}\dot{q}^2 .
\end{align}
The pumping current is $j=\int \frac{dq}{2\pi}\dot{x}$ when the Fermi energy lies at the energy gap. The anomalous velocity from $\partial_q A_{\tau}$ contributes to a quantized pumping current.
The nonadiabatic correction modifies the group velocity through $\delta E_{\rm na}'$, which does not have net contribution to the pumping current. Due to the absence of external fields, $\dot{q}=0$ and thus both the geometric and geodesic velocities from $G_{qq}$ and $\Gamma_{q,qq}=\partial_qG_{qq}/2$ vanish. From this perspective, the nonadiabatic corrections can only modify the pumped current through modifying $A_{\tau}$. However, as the corrections are gauge invariant, the integral of the corresponding Berry curvature over the Brillouin zone vanishes. Thus, the charge pumping remains quantized. 

\subsection{Static Helical Texture}
Here, we study the effect of helical texture of the exchange field, which may arise from spin textures. As an example, we set $m_x=m\cos(kx)$ and $m_y=m\sin(kx)$ with wavelength $2\pi/k$ much longer than the lattice constant. In this case, 
\begin{align}
    \Lambda^{00} =& \,\dot{x}\,A_x - H_1^{00} \\
    \Lambda^{10} =& \, \dot{q}\,A_q^{10} + \dot{x}\,A_x^{10} 
\end{align}
where $A_x=-k(1+\frac{vq}{\Delta})$, $H_1^{00}=vk\frac{g^2m^2}{\Delta^2}$, $A_q^{10}=\frac{i v g m}{2 \Delta^2}$, and $A_x^{10}=-\frac{gmk}{2\Delta}$. As a result, the nonadiabatic effects are only manifested in the nonadiabatic metric, whereas the corrections to Berry connections and energy vanish.
The effective Lagrangian reads:
\begin{align}
    L_{\rm eff} 
    &= - \dot{{q}}  {x} - E_{0}(q) + \dot{x}A_x - H_1^{00} \notag \\
    &+ \frac{1}{2}(G_{qq}\dot{q}^2 + G_{xx}\dot{x}^2)
\end{align}
where $G_{qq}=2\frac{|A_q^{10}|^2}{{\color{black}2}\Delta}$, $G_{xx}=2\frac{|A_x^{10}|^2}{{\color{black}2}\Delta}$, and $G_{qx,xq}=0$ as $A_q^{01}A_x^{10}$ is pure imaginary. In the absence of external fields, the corresponding equation of motion reads
\begin{align}
\dot{{x}} 
    &= \partial_q E - {{F}}_{qx} \dot{{x}} + {G}_{qq} \ddot{{q}} + {\Gamma}_{q,qq}\dot{q}\dot{q}+ {\Gamma}_{q,xx}\dot{x}\dot{x} \\
-\dot{{q}} 
    &= \partial_x E - {{F}}_{xq} \dot{{q}} + {G}_{xx} \ddot{{x}}  + 2{\Gamma}_{x,xq}\dot{x}\dot{q}
\end{align}
where $F_{qx}=\partial_qA_x=-F_{xq}$ as $A_q=0$. In the absence of $G_{qq,xx}$, i.e., the adiabatic limit, the above equations become algebraic equations of $\dot{q}$ and $\dot{x}$ with solutions $\dot{x}_a = \partial_q E/(1+F_{qx})$ and $\dot{q}_a = -\partial_x E/(1+F_{qx})$. Then one can solve the $(\dot{x},\dot{q})$ iteratively by taking the adiabatic solutions as the zeroth-order solution. 

Alternatively, one can reexpress the above equation as forced geodesic equation as shown in Eq.~\eqref{eq:analogousgravity} in phase space spanned by $x$ and $q$. This space is equipped with a nonadiabatic metric $G$ that is diagonal and invertible in this case. Solving this equation now requires initial values of both $(x,q)$ and $(\dot{x},\dot{q})$. It is noted that, while the equation mimics the dynamics of quasiparticle dynamics in a gravitational field defined by $G$, the effective magnetic field from the symplectic matrix is the dominant force on the particle dynamics. This makes it different from the studies of particle dynamics on curved surfaces or deformed lattice~\cite{dong2018geometrodynamics, li2023geodynamics, jiang2022geometric}.

\subsection{Static Collinear Density Wave}
We consider slowly varying $m_x$ in space and $m_y=0$. In contrast to the case with helical texture, the energy gap here, $2\Delta$, is spatially varying. Furthermore, the collinear structure of $\bm{m}$ leads to a zero $\Lambda^{00}$, indicating that the nonadiabatic effects are the leading-order correction to the wave packet dynamics. The corresponding interband Berry connections and gradient correction are $A_q^{10} = \frac{i v g m}{2 \Delta^2}$, $A_x^{10} = \frac{- i v qg}{2 \Delta^2}\partial_x{m}_x$, and $H_1^{10} = i \frac{v g^3 m^2}{2 \Delta^3} \partial_x{m}_x$
where $m=|m_x|$. In this case, nonzero $A_q^{10}$ and $A_x^{10}$ lead to a nonzero nonadiabatic metric, which however is singular, i.e., $\det(G)=0$. 
Thus, the equations of motion cannot be converted into the forced geodesic equation of motion in phase space. $H_1^{10}$ contributes to a correction of the wave packet energy, which is in the second order of the gradient.

The cross terms of $H_1^{10}$ and $A_{q,x}^{10}$ give rise to nonzero corrections to the Berry connections. The correction to momentum-space Berry connection is $\delta A_q = -\frac{v^2m^3g^4}{2\Delta^6}\partial_x m_x$. This is an even function of $q$, indicating a nonzero electric polarization induced by the gradient correction: $P=\int \frac{dq}{2\pi}\delta A_q$. This differs from the helical spin texture case where the local polarization is always zero. The correction to real-space Berry connection is $\delta A_x = \frac{qv^2m^2g^4}{2\Delta^6}(\partial_x m_x)^2$. This $\delta A_x$ plays the role of an effective vector potential, which is an odd function of $q$. The nonadiabatic Berry connections in $q$ and $x$ space modify the electronic wave packet dynamics as well as the density of states. 

\section{Summary}
We introduced the leading-order nonadiabatic corrections to wave-packet dynamics in crystalline materials by deriving an effective Lagrangian valid under slowly varying spatial and temporal perturbations. These corrections arise from two sources: (i) interband coupling induced by the dynamical evolution of the wave-packet centers $(\bm{q}_c, \bm{x}_c)$, and (ii) spatial gradient and temporal variation of the local Hamiltonian. The former yields a nonadiabatic metric, a metric tensor on the phase space spanned by $(\bm{q}_c, \bm{x}_c)$, closely related to the energy-gap-renormalized quantum metric. This reformulates wave-packet motion as forced geodesic dynamics in a curved space. The latter produces an energy correction, which in turn modifies the group velocity and external forces. Cross terms between these effects yield corrections to the geometric Berry connections, altering the anomalous velocity and generating emergent electromagnetic fields from Hamiltonian variations. As an application, we analyzed 1D Dirac electrons in a slowly varying exchange field $\bm{m}$. Whereas adiabatic dynamics are governed by directional variations of $\bm{m}$ induced by, e.g., spin textures or spin precession, which generate emergent gauge fields, our results demonstrate that longitudinal variations in the magnitude of $\bm{m}$ are equally crucial in the nonadiabatic regime, leading to the emergence of the nonadiabatic metric, corrections to Berry connections, and energy shifts.

\textbf{\textit{Note Added---}} Upon completing this work, we became aware of two related studies~\cite{onishi2025emergent, yoshida2025emergent}. The first investigates the dynamics of a two-dimensional electron gas with quadratic dispersion subjected to a spatially varying spin texture using a Hamiltonian formalism~\cite{onishi2025emergent}. The second also studies the electron gas. While this work used the Lagrangian formalism~\cite{yoshida2025emergent}, it does not capture the gradient corrections addressed in our study. The corrections of the nonadiabatic effects to the density of states in phase space have also been studied recently~\cite{mameda2025quantum, maranzana2026semiclassical}.

\begin{acknowledgments}
This research was supported by the US National Science Foundation (NSF) through the University of Delaware Materials Research Science and Engineering Center, DMR-2011824. 
\end{acknowledgments}

%

\end{document}